\documentclass[letterpaper, twocolumn, oneside]{article}
\usepackage{usenix}
\usepackage{hyperref}
\PassOptionsToPackage{hyphens}{url}\usepackage{hyperref}
\usepackage{siunitx}
\usepackage{graphicx}
\usepackage{comment}
\usepackage{mdwlist}
\usepackage[small]{caption}
\usepackage{pifont}
\usepackage{color}
\usepackage{times}
\usepackage[sort&compress,square,numbers]{natbib}
\usepackage{pdfpages}
\usepackage{balance}
\usepackage{amsmath}
\usepackage{amssymb}
\usepackage{xspace}
\usepackage{wrapfig}
\usepackage{pifont}
\usepackage{booktabs}
\usepackage{paralist}
\usepackage{titlesec}
\usepackage{usenix}
\usepackage{hyperref}

\usepackage{titling}
\usepackage{amsmath}
\usepackage{hyperref}

\begin{document}
\setlength{\droptitle}{-4em}   
\title{Defensive Distillation is Not Robust to Adversarial Examples}
%
\date{}
\author{Nicholas Carlini and David Wagner \\ University of California, Berkeley}
\maketitle

\section*{Abstract}
We show that defensive distillation is not secure:
it is no more resistant to targeted misclassification attacks
than unprotected neural networks.


\section{Introduction}

It is an open question how to
train neural networks so they will be robust to adversarial examples
\cite{szegedy2013intriguing}.
Defensive distillation \cite{distillation} was recently proposed as an approach
to make
feed-forward neural networks robust against adversarial examples.

In this short paper, we demonstrate that defensive distillation is not
effective.
We show that, with a slight modification to a standard attack,
one can find adversarial examples on defensively distilled  networks.
We demonstrate the attack on the MNIST \cite{mnist} digit recognition task.

Distillation prevents existing
techniques from finding adversarial examples by increasing the magnitude of
the inputs to the softmax layer.
This makes an unmodified attack fail.
We show that if we artificially reduce the magnitude of the input
to the softmax function, and make two other minor changes, the attack succeeds.
Our attack achieves successful targeted misclassification on $96.4\%$ of
images by changing on average $4.7\%$ of pixels.



\section{Background}

\subsection{Neural Networks and Notation}

We assume familiarity with neural networks \cite{lecun1998gradient}, adversarial examples
\cite{szegedy2013intriguing},
and defensive distillation \cite{distillation}. We briefly review the key details and notation.

Let $F(\theta, x) = y$ be a neural network with model parameters $\theta$ evaluated on
input instance $x$, with the last layer a softmax activation. Call the second-to-last layer (the
layer before the
the softmax layer) $Z$, so that $F(\theta, x) = \text{softmax}(Z(\theta, x))$. When $F$
is used for classification tasks, each output $y_i$ corresponds to the
predicted probability that the object $x$ is labelled as class $i$. We let
$C(\theta, x) = \arg \max_i \ F(\theta, x_i)$ correspond to the classification
of $x$. Often we ommit $\theta$ for clarity. In this paper we are concerned with neural
networks used to classify greyscale images.

\bigskip \noindent
\textbf{Adversarial examples} \cite{szegedy2013intriguing}
are instances $x'$ which are very close to a
valid instance $x$ with respect to some distance metric,
but where $C(\theta, x) \ne C(\theta, x')$.
Given an input image $x$ and a target class $t$
(different than the correct classification of $x$),
a \emph{targeted} misclassification attack is possible if an
adversary can find an adversarial example $x'$ such that $C(\theta, x') = t$
and $x'$ is very similar to $x$. As the targetted attack is more powerful we
focus on this.


\bigskip \noindent
\textbf{Papernot's attack} \cite{papernot2016limitations}
is an algorithm for finding adversarial examples that are close to the
original image with respect to the $L_0$ distance metric
(i.e., few pixels are changed).
We describe the attack as proposed by Papernot \emph{et al}.
The reader is referred to their paper for motivation and explanation
of why it succeeds~\cite{papernot2016limitations}.

The attack consists of many iterations of a greedy selection procedure.
In each iteration, Papernot's attack chooses pixels $(p^*, q^*)$
to change that will make the desired target classification $t$ most likely.
It sets these pixels to either fully-on or fully-off
to make $t$ the most likely.
This is repeated until either (a) the image is classified
as the target, or (b) more than 112 pixels are changed (the threshold
determined to be detectable).

In order to pick the best pair of pixels to modify, the attack uses the
gradient of the network to approximate their importance. Let
\begin{align*}
  \alpha_{pq} & = \sum\limits_{i \in \{p,q\}} \frac{\partial Z(x)_t}{\partial x_i} \\
  \beta_{pq} & = \left(\sum\limits_{i \in \{p,q\}} \sum\limits_{j=0}^9 \frac{\partial Z(x)_j}{\partial x_i}\right)-\alpha_{pq}
\end{align*}
so that $\alpha_{pq}$ represents how much changing $(p, q)$ will change the target
classification, and $\beta_{pq}$ represents how much changing $(p, q)$ will change
all other outputs. Then the algorithm picks
\begin{align*}
  (p^*, q^*) & = \arg\max\limits_{(p, q)}\  (-\alpha_{pq} \cdot \beta_{pq}) \cdot (\alpha_{pq} > 0) \cdot (\beta_{pq} < 0)
\end{align*}
so that $\alpha>0$ (the target class
is more likely), $\beta<0$ (the other classes become less likely),
and $-\alpha \cdot \beta$ is largest.
Notice that Papernot's attack uses the output of the second-to-last
layer $Z$, the logits, in the calculation of the gradient: the
output of the softmax $F$ is \emph{not} used.\footnote{The authors indicated via
  personal communication that they use $F$, the output of the softmax, when attacking
  defensively distilled networks. This is different than the attack as initially
  presented. This does not change any of our results.}

\bigskip \noindent
\textbf{Defensive distillation} was proposed to prevent adversarial
examples \cite{distillation}. It is trained in three steps:
\begin{enumerate}
\item Train a network (the \emph{teacher}) using standard techniques.
  In this network, the output is given by
  $F(\theta, x) = \text{softmax}(Z(\theta, x)/T)$ for some temperature $T$.
  As $T \to \infty$ the distribution approaches uniform; as $T \to 0^+$
  the distribution approaches the hard maximum; standard softmax uses $T=1$.
\item Evaluate the teacher network on each instance of the training set to
  produce \emph{soft labels}. These soft labels contain additional information;
  for example the network may say a digit $x$ has a $80\%$ chance of being a 7
  and a $20\%$ chance of being a 1.
\item Train a second network (the \emph{distilled network}) on the soft
  labels again using temperature $T$. By training on the soft labels,
  the model should overfit the data less and try to be more regular.
\end{enumerate}
Finally, to classify an input, run the distilled network using temperature $T=1$.
By training at temperature $T$, the logits (the inputs to the softmax)
become on average $T$ times larger in absolute value to minimize the cross-entropy loss.
This causes the network to become significantly more confident in its predictions
when evaluating on temperature $1$.

We choose $T=100$, which was found to be the most difficult to attack of
the temperatures that were proposed for use with defensive
distillation~\cite{distillation}.
Defensive distillation with $T=100$
lowers the success probability of Papernot's attack to $0.45\%$
and increases the average number of pixels required to change
the classification from $2\%$ to $14\%$.

\subsection{Our Implementation}
We use TensorFlow \cite{tensorflow} to re-implement defensive distillation
and our variant on Papernot's attack.

We use the same 9-layer network architecture as proposed by
Papernot \emph{et al.} \cite{distillation}.  We use a slightly smaller
learning rate, which we found to converge more quickly.
We train on the MNIST \cite{mnist} data set.
Our baseline model achieves $99.4\%$ accuracy;
the distilled network $99.1\%$. This is comparable to the state-of-the-art.

Our model creation and attack code is open source and available at
\url{http://nicholas.carlini.com/code/nn_defensive_distillation};
all of the data in this paper is reproducible from the provided code.

\section{Breaking Distillation}





We demonstrate that defensive distillation is not effective by modifying Papernot's $L_0$
attack described above. 

\begin{figure}[t]
\begin{centering}
  \includegraphics[scale=0.45]{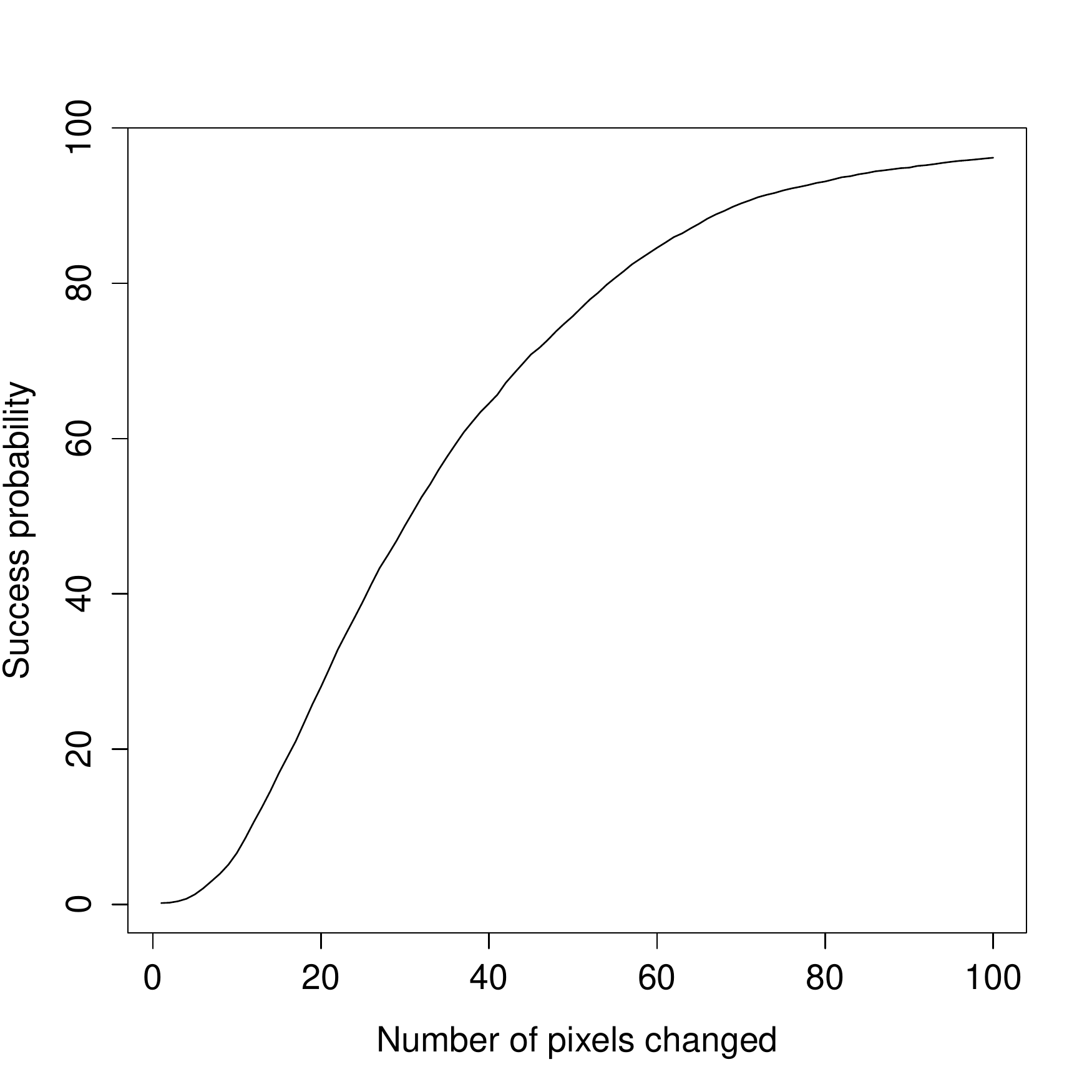}
\end{centering}
\caption{Cumulative density function showing probability that we we can find a
  targetted adversarial example on a defensively distilled network a given
  number of pixels changed. }
\end{figure}

\paragraph{Examining why distillation stops Papernot's attack.}
Recall that distillation as used above does not take the derivative with respect
to the last softmax layer, but instead with the second-to-last layer.
When dealing with the input to the softmax layer (the logits), it is important to realize the differences in relative
impact of terms. If the smallest input to the softmax layer is $-100$, then, after the softmax layer, the corresponding output becomes
practically zero. If this input changes from $-100$ to $-90$, the output will still
be practically zero. However, if the largest input to the softmax layer is $10$, and it changes to $0$,
this will have a massive impact on the softmax output.

Relating this to the $\alpha$ and $\beta$ used above, because Papernot's
attack computes the gradient of the \emph{input} to the softmax layer,
$\alpha$ and $\beta$ represent the size of the change at the input
to the softmax layer.
It is perhaps surprising that Papernot's
attack works on un-distilled networks: it treats all changes as being of
equal importance, regardless of how much they change the softmax output.
For example it will not change a pixel increases $\alpha$
from $10$ to $20$ if it would also increase $\beta$ from $-100$ to $-80$, even if
the latter is merely because some value would change from $-40$ to $-20$.
Thus, Papernot's algorithm may fail to find an adversarial example
even if one exists.

When we train a distilled network at temperature $T$ and then test it at temperature $1$,
we effectively cause the inputs to the softmax to become larger by a factor of $T$.
By minimizing the cross entropy during training, the output of the softmax is forced to
be close to $1.0$ for the correct class and $0.0$ for all others.
Since $Z(\theta,x)$ is
divided by $T$, the network will simply learn to make the $Z(\theta,\cdot)$ values $T$ times larger than
they otherwise would be.
(Positive values are forced to become about $T$ times larger;
negative values are multiplied by a factor of about $T$ and thus become even more negative.)
Experimentally, we verified this fact: the mean value of the $L_1$ norm of
$Z(\theta,x)$ (the logits) on the undistilled
network is $5.8$ with standard deviation $6.4$; on the distilled network (with $T=100$),
the mean is $482$ with standard deviation $457$.
In effect, this magnifies the sub-optimality noted above and causes
Papernot's attack to fail spectacularly when applied to the distilled network.

\paragraph{Modifying the attack.}
Fixing this issue requires only minor modifications to
Papernot's attack. First, instead of taking the gradient
of the inputs to the softmax, we instead take the gradient of the
actual output of the network. However, now the gradients vanish due to the large
absolute value of the inputs to the softmax. To resolve this, we 
artificially divide the inputs to the softmax by $T$ before using them. Let
\begin{equation*}
  \hat{F}(\theta, x) = \text{softmax}(Z(\theta, x)/T)
\end{equation*}
Now the inputs to the softmax are of acceptable size and the gradients no longer vanish.
Second, we are able to achieve slightly
better accuracy by taking the maximum over $\alpha - \beta$ instead of the
product used earlier. In fact, with these modifications, we do
not lower accuracy even if we search over one pixel at a time instead of pairs of pixels,
which is significantly ($768 \times$) more efficient. Thus, when selecting the best
pixel to modify, we select
\begin{equation*}
  p^* = \text{arg}\max\limits_{p}\ 2 \frac{\partial \hat{F}(x)_t}{\partial x_p} - \sum\limits_{j=0}^9 \frac{\partial \hat{F}(x)_j}{\partial x_p}
\end{equation*}
where we have simplified terms.

For the strongest setting of $T=100$, we achieve a successful
targetted misclassification rate of $96.4\%$, changing on average $36.4$
pixels out of $768$ ($4.7\%$ of the pixels).
Figure~1 shows a CDF
of the number of pixels required to change the classification for successful attacks.
We verified that our attack works for any other setting of $T$ from $1$
to $100$, the same range studied initially.

As a baseline for comparison, we ran our modified attack against a standard
network trained without distillation. Note that this is not an entirely fair
comparison: we have made changes to increase success against distilled networks.
Despite this, our attack succeeds $86\%$ of the time with on average $45$ pixels
changed.
This indicates that the network trained with defensive distillation
is no more secure against adversarial examples than a standard network
trained without distillation.




\section{Conclusion}
When creating a defense of any form, it is important to analyze how
it might be attacked.
It is not sufficient to demonstrate that
it defends against existing attacks; it must also be effective
against future attacks.

While it is impossible to test against all possible future attacks,
we encourage designers to look for an argument
that existing attacks can not be adapted.
After observing that an attack fails on a proposed defense, it would be
useful to understand \emph{why} the attack fails. As we have shown in this case study,
it is possible that
the attack only fails due to superficial reasons, and small modifications
can result in failure of the defense.
While Papernot's attack is powerful enough to break
unhardened networks, it makes no claims of optimality, and
demonstrating that a defense successfully stops a sub-optimal attack does
not imply it will stop all other attacks.

Defending against adversarial examples remains a challenging open problem
for the field.
When proposing a defense, we recommend researchers evaluate why the
defense works and whether it will be effective against attacks targeted at
that specific defense.

{\footnotesize
\bibliographystyle{abbrvnat}
\bibliography{paper}
}

\end{document}